\documentclass[aps,prb,twocolumn,superscriptaddress]{revtex4-1}
\usepackage{bm,amsmath}
\usepackage{siunitx}

\usepackage[whole]{bxcjkjatype} 
\usepackage{graphicx}
\usepackage[colorlinks=true,urlcolor=blue,citecolor=blue,linkcolor=blue,breaklinks=true]{hyperref}
\input glyphtounicode
\usepackage{txfonts}
\usepackage{xcolor}

\graphicspath{{figs/}}

\DeclareGraphicsExtensions{.pdf}

\usepackage{wrapfig}
\usepackage{epsfig}
\usepackage{amsfonts}
\usepackage{amssymb}

\begin{document}

\title{
Spin Hall effect in 2D metallic delafossite PtCoO$_2$  and vicinity topology
}

\author{Sota Kitamura}
\affiliation{Max-Planck-Institut f\"ur Physik komplexer Systeme, N\"othnitzer Stra{\ss}e 38, 01187 Dresden, Germany}

\author{Hidetomo Usui}
\affiliation{Department of Physics, Osaka University, 1-1 Machikaneyama-cho, Toyonaka, Osaka, 560-0043, Japan}

\author{Robert-Jan Slager}
\affiliation{Max-Planck-Institut f\"ur Physik komplexer Systeme, N\"othnitzer Stra{\ss}e 38, 01187 Dresden, Germany}

\author{Adrien Bouhon}
\affiliation{Department of Physics and Astronomy, Uppsala University, Box 516, SE-751 21 Uppsala, Sweden}
\affiliation{NORDITA, Roslagstullsbacken 23, 106 91 Stockholm, Sweden}

\author{Veronika Sunko}
\affiliation{Max-Planck-Institut f\"ur Chemische Physik fester Stoffe, N\"othnitzer Stra{\ss}e 40, 01187 Dresden, Germany}
\affiliation{SUPA, School of Physics and Astronomy, University of St. Andrews, St. Andrews KY16 9SS, United Kingdom}

\author{Helge Rosner}
\affiliation{Max-Planck-Institut f\"ur Chemische Physik fester Stoffe, N\"othnitzer Stra{\ss}e 40, 01187 Dresden, Germany}

\author{Philip D. C. King}
\affiliation{SUPA, School of Physics and Astronomy, University of St. Andrews, St. Andrews KY16 9SS, United Kingdom}

\author{Joseph Orenstein}
\affiliation{Department of Physics, University of California, Berkeley, CA}

\author{Roderich Moessner}
\affiliation{Max-Planck-Institut f\"ur Physik komplexer Systeme, N\"othnitzer Stra{\ss}e 38, 01187 Dresden, Germany}

\author{Andrew P. Mackenzie}
\affiliation{Max-Planck-Institut f\"ur Chemische Physik fester Stoffe, N\"othnitzer Stra{\ss}e 40, 01187 Dresden, Germany}
\affiliation{SUPA, School of Physics and Astronomy, University of St. Andrews, St. Andrews KY16 9SS, United Kingdom}

\author{Kazuhiko Kuroki}
\affiliation{Department of Physics, Osaka University, 1-1 Machikaneyama-cho, Toyonaka, Osaka, 560-0043, Japan}

\author{Takashi Oka}
\affiliation{Max-Planck-Institut f\"ur Physik komplexer Systeme, N\"othnitzer Stra{\ss}e 38, 01187 Dresden, Germany}
\affiliation{Max-Planck-Institut f\"ur Chemische Physik fester Stoffe, N\"othnitzer Stra{\ss}e 40, 01187 Dresden, Germany}

\begin{abstract}
The two-dimensional (2D) metal PtCoO$_2$ is renowned for 
the lowest room temperature resistivity among all oxides, close to that of 
the top two materials Ag and Cu~\cite{Shannon1971-3,MackenzieReview}. 
In addition, we theoretically predict a strong intrinsic spin Hall effect~\cite{Murakami1348,Sinova04}. 
This originates from six strongly-tilted Dirac cones 
that we find in the electronic structure near the Fermi surface, 
where a gap is opened by large spin-orbit coupling (SOC).  
This is underpinned by rich topological properties; 
in particular, the phenomenology of a mirror Chern metal~\cite{Teo_mirror_Z2} 
is realized not exactly, but very accurately, on account of 
an \textit{approximate} crystalline symmetry.
We expect that such `vicinity topology' to be a feature of relevance well beyond this material. 
Our Wilson loop analysis indicates further elaborate features such as fragile topology~\cite{Po_Fragile,BouhonSlager_1,Bradlyn_WilC3}. 
These findings highlight PtCoO$_2$ as a promising material for spintronic applications 
as well as a platform to study the interplay of symmetry and topology. 
\end{abstract}

\maketitle

\begin{figure}[t]
\centering 
\includegraphics[width=.9\hsize]{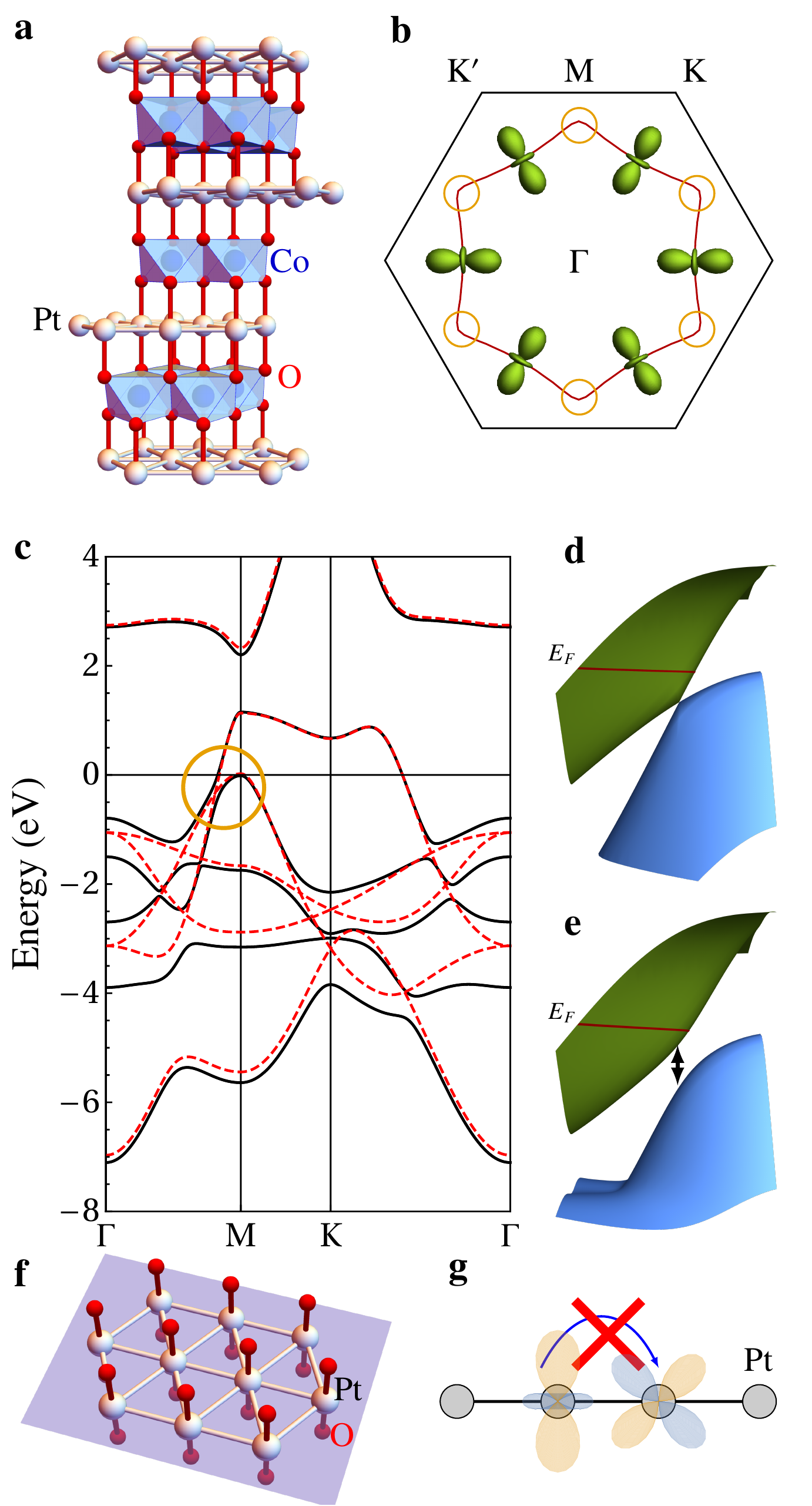}
\caption{
\textbf{Electronic structure.}
\textbf{a}, Crystal structure of PtCoO$_2$.
\textbf{b}, Hexagonally warped Fermi surface (red curve) and the corresponding orbital eigenfunction (green). 
\textbf{c}, Band structure along high symmetry lines. 
Solid and dashed lines show the band dispersions with and without spin-orbit coupling, respectively.
The spin-orbit coupling is $\lambda=\SI{0.55}{\eV}$. 
\textbf{d}, \textbf{e}, 3D picture of the tilted Dirac cone before and after adding the spin-orbit coupling.
Their locations are indicated by circles in b,~c. 
\textbf{f}, Pt layer with adjacent O atoms. 
\textbf{g}, Schematic picture of a typical hopping process forbidden in a basal-mirror symmetric structure. 
}
\label{fig:es}
\end{figure}

PtCoO$_2$ crystallizes in the delafossite structure in which,
as shown in Fig.~\ref{fig:es}a, Pt and Co respectively form triangular lattices and are stacked alternately~\cite{Shannon1971-2}. 
Their formal valences and configuration are Pt$^{1+}$ with $5d^9$ and Co$^{3+}$ with $3d^6$.
Within this ionic-bonding picture, Co forms a band insulating state due to the crystal field splitting,
while Pt is characterized by metallic $d_{3z^2-r^2}$ electrons. 

The conductivity is exceptionally high for an oxide, and the electronic transport is strongly 2D~\cite{Shannon1971-3,Eyert2008}.
The Fermi surface is a single cylinder with hexagonal deformation (Fig.~\ref{fig:es}b), and the 
sharpness seen in the angle resolved photoemission spectroscopy (ARPES) bands indicates that electron
correlation is weak~\cite{Kushwaha2015}.
A naive view of this material is a heterostructure of nearly-free Pt-derived $d_{3z^2-r^2}$ electrons on a triangular lattice 
sandwiched by insulating layers of CoO$_2$~\cite{Seshadri1998,Eyert2008}. 
However, as we show here, PtCoO$_2$ is far from being captured by this picture, but has a rich orbital and topological structure. 
We clarify this by constructing a two-dimensional nine-band tight-binding model on a triangular lattice, 
which consists of Pt $5d$, $6s$ and $6p$ orbitals obtained as 
maximally-localized Wannier functions for the first-principles band structure.

Figure \ref{fig:es}b displays the numerically calculated orbital shape along the Fermi surface, and  we already find unexpected rich features. 
 The orbital is not a simple $d_{3z^2-r^2}$, but lies within 
 the $xy$-plane, e.g. $d_{3x^2-r^2}$-like, and aligns perpendicular to the sides of the hexagon. 
This orbital-momentum locking cannot emerge in the naive single-orbital triangular lattice picture and may 
explain characteristic behaviors of the material such as 
 reduced impurity scattering~\cite{Usui2018}. 

Another remarkable feature is the existence of Dirac cones just below the Fermi level.
This becomes apparent when we switch off the SOC. 
The band structure with and without SOC is shown in Fig.~\ref{fig:es}c,
where we find strongly tilted 2D Dirac cones near
the six corners of the hexagonal Fermi surface (encircled in Fig.~\ref{fig:es}bc and enlarged in \ref{fig:es}d). 
Without SOC, these Dirac cones are robust because of the crystalline symmetry:
A perturbation preserving the crystal symmetry can shift the position of the crossing along the $\Gamma$--M line,
but cannot remove it without merging the six copies at the $\Gamma$ point.
The tilt is large enough that the cone is flipped, such that the electron pockets are absorbed 
into the larger Fermi surface, placing it in the  type II Dirac fermion category~\cite{Soluyanov:cn}.
The SOC of Pt is so strong ($\lambda\sim\SI{0.55}{\eV}$ for $H_\text{SO}=\lambda\bm{L}\cdot\bm{S}$) that 
an energy gap of $\sim\SI{0.43}{\eV}$ opens at the Dirac nodes as shown in Fig.~\ref{fig:es}e, which is consistent with 
recent ARPES observations~\cite{Sunko:td}.
While the large gap washes out the conical energy dispersion,
features of the Dirac nodes are inherited in band topology and transport properties as we detail below.

Before moving to topology, let us summarize the relevant crystalline symmetries. 
The first symmetry is the invariance under 
vertical mirror $\sigma_d:x\rightarrow-x$ (and its two symmetric partners obtained by $C_{3z}$ rotation). 
This protects the six Dirac nodes in the absence of SOC as discussed above.  
The second is inversion symmetry $I:\bm{r}\rightarrow-\bm{r}$. 
In combination with the time-reversal symmetry (TRS), this makes the bands Kramers degenerate.
The third, which plays an important role in the following topological discussion, 
is only approximate, defined by the basal mirror $\sigma_h:z\rightarrow-z$.
While this symmetry is absent in the stacked delafossite structure as a whole, 
it is respected in a single Pt and its adjacent oxygen layers, which determines the electronic structure of the 
2D metallic Pt electrons and oxygen crystal fields (see Fig.~\ref{fig:es}f). 
While we consider a two-dimensional model here, 
we note that orbital functions have three-dimensional structures
and the basal-mirror symmetry $\sigma_h$ imposes a constraint on their hybridization.
For instance, as we depict in Fig.~\ref{fig:es}g, a $d_{xz}$ orbital (odd in $z$) electron cannot 
tunnel to a $d_{3z^2-r^2}$ orbital (even in $z$).
As the mirror operation acts as $i\sigma_z$ for spins, 
SOC $\lambda\bm{L}\cdot\bm{S}$ preserves the mirror symmetry $\sigma_h$.

\begin{figure*}[t]
\centering 
\includegraphics[width=\hsize]{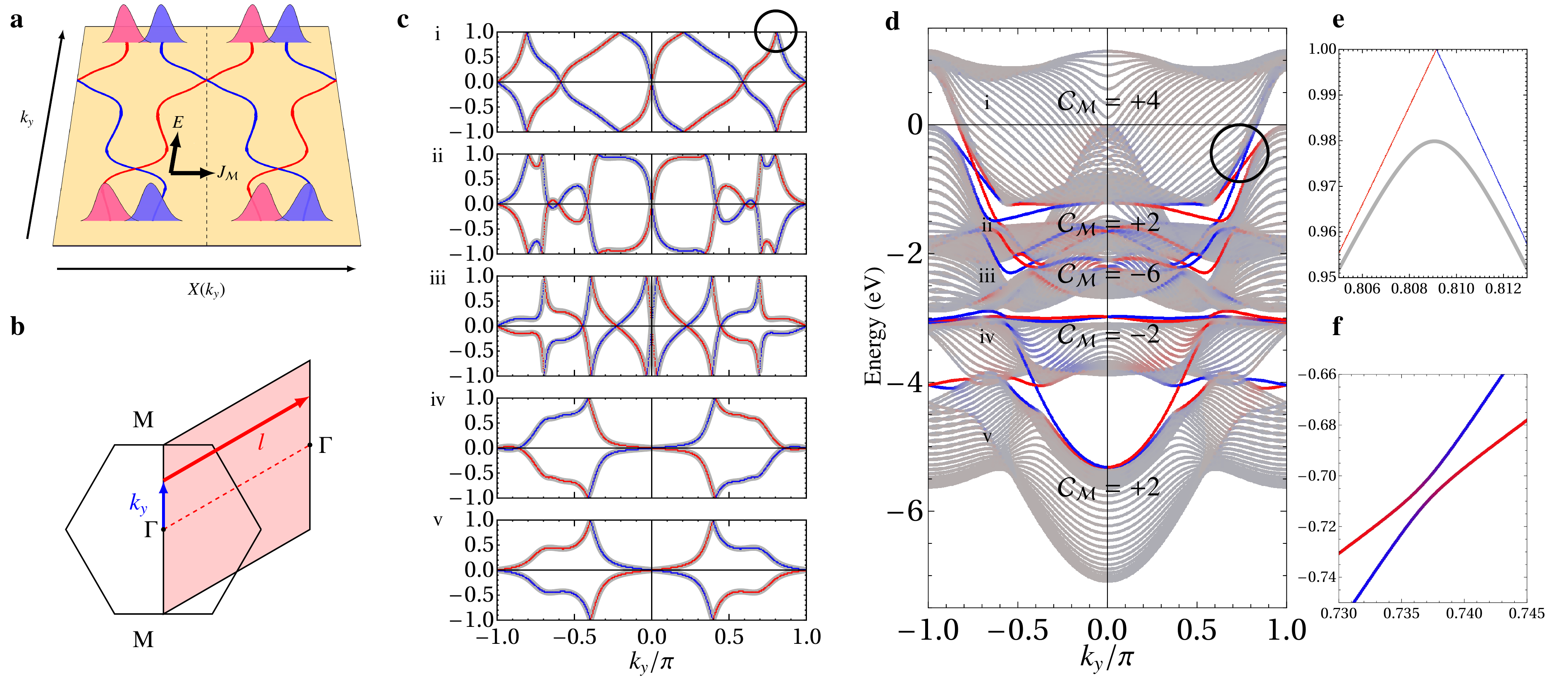}
\caption{
\textbf{Topological structure.}
\textbf{a}, Schematic motion of the Wannier centers as a function of momentum $k_y$.
The red and blue lines corresponds to the trajectory $X(k_y)$ of the two Kramers-degenerate bands (plotted in c). 
$E$ and $J_{\mathcal{M}}$ denote the applied static electric field and the induced mirror Hall current, respectively. 
\textbf{b}, Definition of the momentum path $l$ for the Wilson loop operator $\mathcal{W}[l]$ [equation (\ref{eq:WL})].
\textbf{c}, Flow of Wilson loop eigenvalues $\pm X(k_y)$ (in unit of $a/2$) for each Kramers-degenerate band descending from the Fermi energy.
Red and blue curves are for the mirror-symmetrized model $H_\mathcal{M}$ (for $\sigma_h=+i$ and $-i$ sectors, respectively), and gray is for the full model $H_\text{tot}$.
\textbf{d}, Band structure in a slab geometry where the bulk states are denoted by gray dots, 
and (right) edge states by red and blue dots. Left-edge states are degenerate with right-edge ones.
The color coding of red to blue indicates the mirror expectation value from $\langle\sigma_h\rangle/i=+1$ to $-1$.  
$C_\mathcal{M}$ indicates the corresponding mirror Chern number for each bulk band of the symmetric parent $H_{\mathcal{M}}$.
\textbf{e}, Enlarged Wilson loop eigenvalue flow around the region encircled in c. 
\textbf{f}, Enlarged energy spectrum around the region encircled in d.
}
\label{fig:topo}
\end{figure*}
Now, let us discuss how the present material 
fits inside the theory of band topology classification~\cite{PhysRevLett.95.146802,
FuKane_inversion,Teo_mirror_Z2, Top1, Top2, 
TQC,Poclas,Po_Fragile,Cano_EBRs,BouhonSlager_1,Bradlyn_WilC3}.  
An important role is played by the hybrid Wannier function, a real-space wave function 
constructed from Bloch states localized in one direction but retaining plane wave character in other directions~\cite{PhysRevB.83.035108}. 
The motion of the Wannier center in crystals can be visualized using the 
non-Abelian Wilson loop~\cite{PhysRevB.84.075119,PhysRevB.83.035108} defined as ($\mathcal{P}$: path ordered product)
\begin{gather}
\mathcal{W}[l]=\mathcal{P}\exp\left[-i\int_{l}d\bm{k}\cdot\bm{\mathcal{A}}(\bm{k})\right],\label{eq:WL}\\
\mathcal{A}_\mu^{mn}(\bm{k})=-i\langle\bm{k}m|\partial_{k^\mu}|\bm{k}n\rangle,
\end{gather}
where $|\bm{k}n\rangle$ is the cell-periodic Bloch wave function of the $n$-th band.
$m,n$ run over 
the (arbitrary) subset that constitutes the hybrid Wannier function of interest, 
and here, we select each Kramers pair as the subset. 
The eigenvalues of 
$\mathcal{W}[l]$ are expressed as $e^{\pm2\pi iX(k_y)/a}$
for a loop $l$ wrapping the Brillouin zone and parametrized by $k_y$ 
as in Fig.~\ref{fig:topo}b ($a$: unit cell length).
We can investigate the band topology 
using the Wannier-center positions  $\pm X(k_y)\in (-a/2,a/2]$ of the Kramers pair. 
The red and blue lines in Fig.~\ref{fig:topo}a display schematically their motions as a function of momentum $k_y$. 
A non-trivial topology is indicated if the Wannier centers move to a different unit cell as $k_y$ is increased by $2\pi$. 
One can intuitively understand this as a Hall effect. 
When a dc-electric field is applied to the crystal in the $y$-direction, the 
momentum is replaced by $k_y-\frac{e}{c}Et$ within the Peierls approximation ($E$: field strength, $t$: time). 
This leads to a Hall current $J_M$ as the Wannier-centers move perpendicular to the electric field. 
In contrast to the usual Hall effect, this Hall current is charge neutral because the 
two states in the Kramers pair move in opposite directions. 

The present material PtCoO$_2$ turns out to be \textit{trivial} 
within the 2D classification scheme of band topology, which is due to the 
weak violation of basal mirror symmetry $\sigma_h$ (We discuss the 3D classification in Supplementary~\ref{sec:3D} in which 
we show that an intersection of the CoO$_2$ bands above the Fermi level makes the Fermi surface band $\mathbb{Z}_2$ non-trivial).
However, this does not mean that the physical properties of the material are trivial. 
On the contrary, we show that this material realizes the physics of the mirror Chern metal~\cite{Teo_mirror_Z2},
i.e., the bands and edge states are approximately quantified by the mirror Chern number. 
In order to clarify this, let us decompose the Hamiltonian into $H_\text{tot}=H_{\mathcal{M}}+H_{\overline{\mathcal{M}}}$
to consider the basal mirror-symmetric part $H_{\mathcal{M}}$ first, and then discuss the perturbative effect of 
the anti-symmetric part $H_{\overline{\mathcal{M}}}$ later.
The motion of the two Wannier centers $\pm X(k_y)$ associated with each Kramers pair of bands are plotted in Fig.~\ref{fig:topo}c,
where red-blue and gray symbols are the results for $H_{\mathcal{M}}$ and $H_\text{tot}$, respectively.
We see that the trajectories for $H_{\mathcal{M}}$ wrap the unit cell several times, and the winding number can be 
associated with mirror Chern numbers. 
Since mirror symmetry $\sigma_h$ is present for $H_{\mathcal{M}}$, we can split the Kramers-degenerate bands into two mirror 
sectors $M_\pm=\{|\mu_\pm\rangle;\;\sigma_h|\mu_\pm\rangle=\pm i|\mu_\pm\rangle\}$, where
each band has its own Chern number $\mathcal{C}(M_\pm)$. 
While the total Chern number $\mathcal{C}(M_+)+\mathcal{C}(M_-)$ vanishes due to TRS,
the difference, the mirror Chern number
$\mathcal{C}_{\mathcal M}=(\mathcal{C}(M_+)-\mathcal{C}(M_-))/2$
is a topological quantity~\cite{Teo_mirror_Z2}
coinciding with the wrapping of the Wannier center resulting in
$\mathcal{C}_{\mathcal M}= +4,+2,-6,-2,+2$ for bands descending from the Fermi energy. 
The effect of the mirror-breaking term $H_{\overline{\mathcal{M}}}$ turns out to be 
small as the gray and red-blue symbols appear to be on top of each other. 
However, as shown in Fig.~\ref{fig:topo}e, small gaps open at crossing points of the Wannier centers 
except for those at $k_y=0,\pi$ that are protected by Kramers degeneracy. 
As a consequence of the approximate mirror symmetry,
pronounced helical edge states appear in the spectrum (Fig.~\ref{fig:topo}d). 
The number of edge states appearing between the bands 
is given by the difference of the mirror Chern numbers. 
This clearly shows that the present system is characterized by the mirror Chern number,
although the mirror is only an approximate symmetry and edge states have small gaps (see Fig.~\ref{fig:topo}f).
This leads to a concept of \textit{vicinity topology} in which the phenomenology of a topological material 
such as exotic quantum transport or edge properties may become realized even when 
the associated symmetry is not present in the whole structure.

\begin{figure}[b]
\centering 
\includegraphics[width=\hsize]{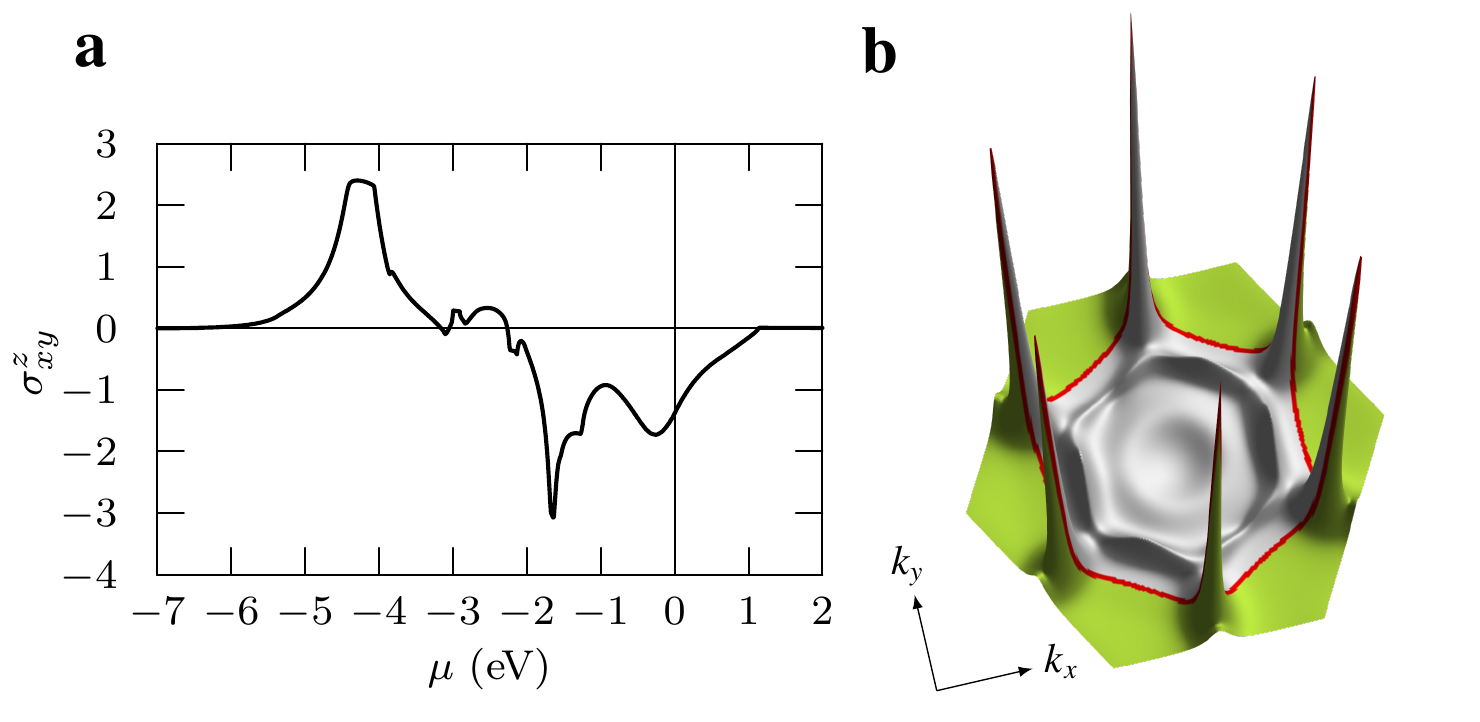}
\caption{
\textbf{Spin Hall effect.}
\textbf{a}, Spin Hall conductivity of PtCoO$_2$ per layer (in unit of the quantized value $e/2\pi$), 
as a function of chemical potential $\mu$.
\textbf{b}, Momentum-resolved spin Hall conductivity of the band forming the Fermi surface. 
Red curve represent the Fermi surface.
}
\label{fig:she}
\end{figure}

As a physical consequence of the vicinity topology, 
we predict that an intrinsic spin Hall effect~\cite{Murakami1348,Sinova04} 
underpinned by the mirror Chern number takes place in PtCoO$_2$. 
Indeed, we find large spin Hall conductivity $\sigma_{xy}^z$ (per layer) as shown in Fig.~\ref{fig:she}a,
which is calculated with the Kubo formula~\cite{PhysRevLett.100.096401}
\begin{gather}
\sigma^z_{xy}=\frac{e}{\hbar}\sum_{n\bm{k}}\Omega^z_n(\bm{k})f_\text{FD}(\varepsilon_{\bm{k}n}-\mu),
\label{eq:SHE1}\\
\Omega^z_n(\bm{k})=\sum_{m\ne n}\frac{2\text{Im}\left[
\langle \bm{k}n|j^z_x|\bm{k}m\rangle\langle \bm{k}m|v_y|\bm{k}n\rangle
\right]}{(\varepsilon_{\bm{k}n}-\varepsilon_{\bm{k}m})^2}.
\label{eq:SHE2}
\end{gather}
Here $\bm{v}$ is the velocity operator, $j^z_x=(1/2)\{\hbar\sigma_z/2,v_x\}$ is the spin current operator, and $f_\text{FD}$ is the Fermi-Dirac function.  
It is convenient to use a hole picture 
starting from a fully-occupied Fermi-surface band. 
The fully occupied state (setting $\mu\sim1$~eV) has no Hall response,
and as we lower the chemical potential $\mu$, the spin Hall conductivity drastically increases. 
Figure~\ref{fig:she}b shows the momentum-resolved spin Hall conductivity,
where the dominant contribution in (\ref{eq:SHE1}) comes from the green region. 
This spin Hall effect can be traced back to the non-zero mirror Chern number in terms of vicinity topology,
as the associated mirror Hall current is well spin polarized; see the spin component of (approximate) mirror eigenstates 
along the Fermi surface, shown in Supplementary~Fig.~\ref{fig:suppl:EY}.

It is interesting to compare this value with the fcc Pt known for 
a large intrinsic spin Hall effect~\cite{PhysRevLett.99.226604}.
The spin Hall conductivity of PtCoO$_2$ shown above translates into $\sim450(\hbar/e)\si{(\ohm\cm)^{-1}}$
for the bulk, while the measured value for fcc Pt 
is~\cite{PhysRevLett.99.226604} $\sim 330 (\hbar/e)\si{(\ohm\cm)^{-1}}$.
The calculated spin Hall coefficient in PtCoO$_2$ is not sensitive to temperature and is expected to persist to room temperature. 
This is because the Fermi surface band is well separated from the lower occupied bands 
and the inter-band cancellation in the integral (\ref{eq:SHE1}) is weak. 

For spintronic applications, it is important for the material to have a long spin relaxation time so that the spin current is well preserved~\cite{Zutic2004}. 
With the extremely long mean free path, 
the present material is already advantageous in this perspective~\cite{Usui2018}. 
However, the richness of the orbital structure and the crystalline symmetry makes it even more attractive. 
Since the present system is inversion symmetric, spin relaxation from spin precession due to 
Rashba or Dresselhaus term, known as the D'yakonov-Perel mechanism~\cite{Dyakonov1972,Zutic2004} is absent. 
Thus,  spin relaxation mainly occurs through the Elliott-Yafet mechanism~\cite{Elliott1954,Yafet1963,Zutic2004}, i.e., 
accompanied by impurity and phonon scattering. 
As detailed in Supplementary \ref{sec:Elliott-Yafet}, the effective mirror symmetry protects the 
mirror Hall current against mirror-preserving scattering processes, and makes the spin relaxation time even longer.  

As a future perspective, we point out that PtCoO$_2$ may serve as an experimental  
testbed for crystalline stable and metastable (or fragile) topologies~\cite{BouhonSlager_1,Po_Fragile,Bradlyn_WilC3}.  
As opposed to stable topology, metastable topology entails topological notions of individual bands that do not necessarily 
persist when many bands are considered. 
The mirror symmetric parent $H_{\mathcal{M}}$ has an additional $C_{2z}=I\sigma_h$ symmetry by virtue of inversion symmetry. 
While the mirror Chern number, protected by $\sigma_h$, is a stable topological indicator, 
the combined anti-unitary $C_{2z}\mathcal{T}$ symmetry protects a metastable topological indicator
also observable in the flow of Wannier centers. 
$C_{2z}\mathcal{T}$ protects the Wannier center crossings at $0$ and $\pm \pi$~within each isolated Kramers pair of bands\cite{Bohm_SW_class,BouhonSlager_1,Bradlyn_WilC3}, 
leading to a metastable $\mathbb{Z}$ classification of the flow of Wannier centers~\cite{BouhonSlager_1} (see Supplementary \ref{Supp:topology} and \ref{sec:fragile} for details). 
What's interesting about  PtCoO$_2$  is that the two topological indicators 
can be separately broken, for example at the top surface. 
If we have a Pt layer surface, $\sigma_h$ is strongly broken since the oxygen is missing on one side, 
while $C_{2z}$ is only weakly broken due to the Co layer coupling as in the bulk. 
Thus, surface sensitive probes such as a scanning tunneling microscope may be able to access 
a state where only metastable topology is realized, making it possible study various theoretical proposals~\cite{BouhonSlager_1,Po_Fragile,Bradlyn_WilC3}.  

To summarize, we revealed here the rich orbital and topological physics behind PtCoO$_2$,
and its interplay with the crystalline symmetry, including approximate ones. 
The retrieved mirror Chern numbers importantly pave the way for experimental and theoretical investigations into \textit{vicinity topologies} that arise due to approximate symmetries, 
falling outside the scope  of recent classification procedures~\cite{Top2, TQC, Poclas} 
but connecting to their physics (edge states, transport etc.) in an effective manner. 
The Wilson loop analysis intimately fits within this analysis as it can be utilized unbiasedly, 
signaling topologies by protecting crossings that can be slightly gapped in case of approximate symmetries.

\section*{Methods}

\subsection*{First-principles band calculation}
We perform a first principles band calculation for PtCoO$_2$ 
adopting the lattice structure obtained in Ref.~\onlinecite{Eyert2008}, and
using the full potential linearized augmented plane wave method 
with the PBE-GGA exchange-correlation functional\cite{PBE}
as implemented in the WIEN2K package\cite{Wien2k}. 
The scalar relativistic approximation is adopted for the relativistic effects of valence electrons.
The value of $RK_\text{max}$ is set to 8, and 1,000 $k$-points are taken for the self-consistent calculation. 
From this band calculation, we obtain maximally localized
Wannier orbitals, which consists of Pt $s\time 1$, $p\times 3$, $d\times 5$ orbitals, using Wannier90\cite{Wannier,Wannier-2,Wannier-3} and Wien2Wannier\cite{w2w} packages.

The spin-orbit coupling, dropped in the scalar relativistic approximation, is explicitly added to the obtained tight-binding model as an on-site term.
The strength of the spin-orbit coupling $\lambda$ is determined 
in such a way as to reproduce the energy level splitting at the $\Gamma$ point in the fully-relativistic band calculation. 
Our estimation $\lambda\sim\SI{0.55}{\eV}$ agrees quite well with 
the theoretical values for the Pt atom calculated in more sophisticated frameworks~\cite{Montalti2006,Kota2012}.

\subsection*{Wilson loop eigenvalues}
To carry out the computation of the Wilson loop operator (\ref{eq:WL}), 
we consider an equidistant $k$-mesh $\{\bm{k}_i\}_{i=1}^L$ on the path $l$ ($L$ is set to 150) and 
an operator $U_i=\langle\bm{k}_im|\bm{k}_{i+1}n\rangle$,
which converges to $\exp[i\bm{\mathcal{A}}(\bm{k}_i)\cdot d\bm{k}]$ in the continuous limit $L\rightarrow\infty$.
The Wilson loop operator $\mathcal{W}[l]$ is obtained as a matrix product $(\prod_{i=1}^L U_i)^\ast$.

\subsection*{Edge states}
The band dispersion shown in Fig.~\ref{fig:topo}d is obtained in a slab geometry, 
i.e., we consider Pt sites at $\bm{R}=(m-n/2,\sqrt{3}n/2)$ (in unit of $a$), 
where the translational invariance is present for one direction, $n\in\mathbb{N}$, 
while the array of atoms is terminated in the other direction, $m=1,\dots,M$ ($M$ is set to 40). 

\begin{acknowledgements}  
We deeply acknowledge discussions with Dr. Masayuki Ochi, Dr. Seunghyun Khim and  Dr. Deepa Kasinathan. 
We thank financial supports from 
ImPACT Program of Council for Science, 
Technology and Innovation, Cabinet Office, Government of Japan (Grant No. 2015-PM12-05-01),
Deutsche Forschungsgemeinschaft (grant SFB 1143),
European Research Council (Grant No. ERC-714193-QUESTDO) and the Royal Society.
J.O. acknowledges support from the Director, Office of Science, Office of Basic Energy Sciences, Materials Sciences and Engineering 
Division, of the U.S. Department of Energy under Contract No. DE-AC02-05CH11231 and the Gordon and Betty Moore Foundation's EPiQS Initiative 
through Grant GBMF4537.
\end{acknowledgements}

\bibliographystyle{naturemag}
\bibliography{reference}

\newpage

\begin{widetext}

\appendix
\setcounter{secnumdepth}{1}
\renewcommand{\figurename}{Supplementary Figure}
\renewcommand{\thesection}{\Alph{section}}
\renewcommand{\theequation}{S\arabic{equation}}
\setcounter{figure}{0}
\setcounter{equation}{0}

\section{Three dimensional classification}
\label{sec:3D}

In the present work, in order to capture in-plane transport properties, 
we have focused on the two-dimensional tight-binding model composed of Pt electrons only.
However, there are also the bands derived from the CoO$_2$ layers in reality, 
and they weakly hybridize with the Pt bands via the inter-layer hopping.
While it should not change the 2D transport properties, 
topological classification of bands may change when Pt-layer bands intersect CoO$_2$-layer bands.
Indeed, the Pt Fermi-surface band intersects the Co-derived bands and 
has a different topological characterization in the three-dimensional classification.
To verify this, we additionally take account of $3d$ orbital of Co atoms and 
$2p$ orbital of O atoms (at two distinct positions), and construct 
a twenty-band 3D tight-binding model.

We show the band structure in Supplementary~Fig.~\ref{fig:suppl:3d}, where
we colored the band by the weights of the additional atoms (blue for Co and red for O).
As in the ionic-bond picture, the Co $3d$ bands split into 
lower three ($a_{1g}$ and $e_g^\pi$) and upper two ($e_g^\sigma$) bands 
due to the crystal field, which results in formation of a band insulator.
We note that the size of this splitting may be underestimated~\cite{Kushwaha2015} 
since the local-density approximation for the density functional fails to capture the correlated nature of $3d$ electrons.
Notably, the Fermi-surface band has intersections with the Co $e_g^\sigma$ bands, and 
undergoes the band inversion at the M point 
leading to an interchange of the inversion eigenvalues.
As a result, the Fermi-surface band has a nontrivial strong index 
in terms of the three-dimensional classification of the topological insulators,
which ensures that the helical edge states appear in between the Fermi-surface band and 
the Co $e_g^\sigma$ bands.

Although this reflects the topological richness of the present system,
we emphasize that the two-dimensional characterization along with the approximate mirror symmetry
given in the main text should be essential for the (in-plane) transport properties,
since the corresponding edge states are responsible for the spin Hall response.

\begin{figure*}[b]
\centering 
\includegraphics[width=0.5\hsize]{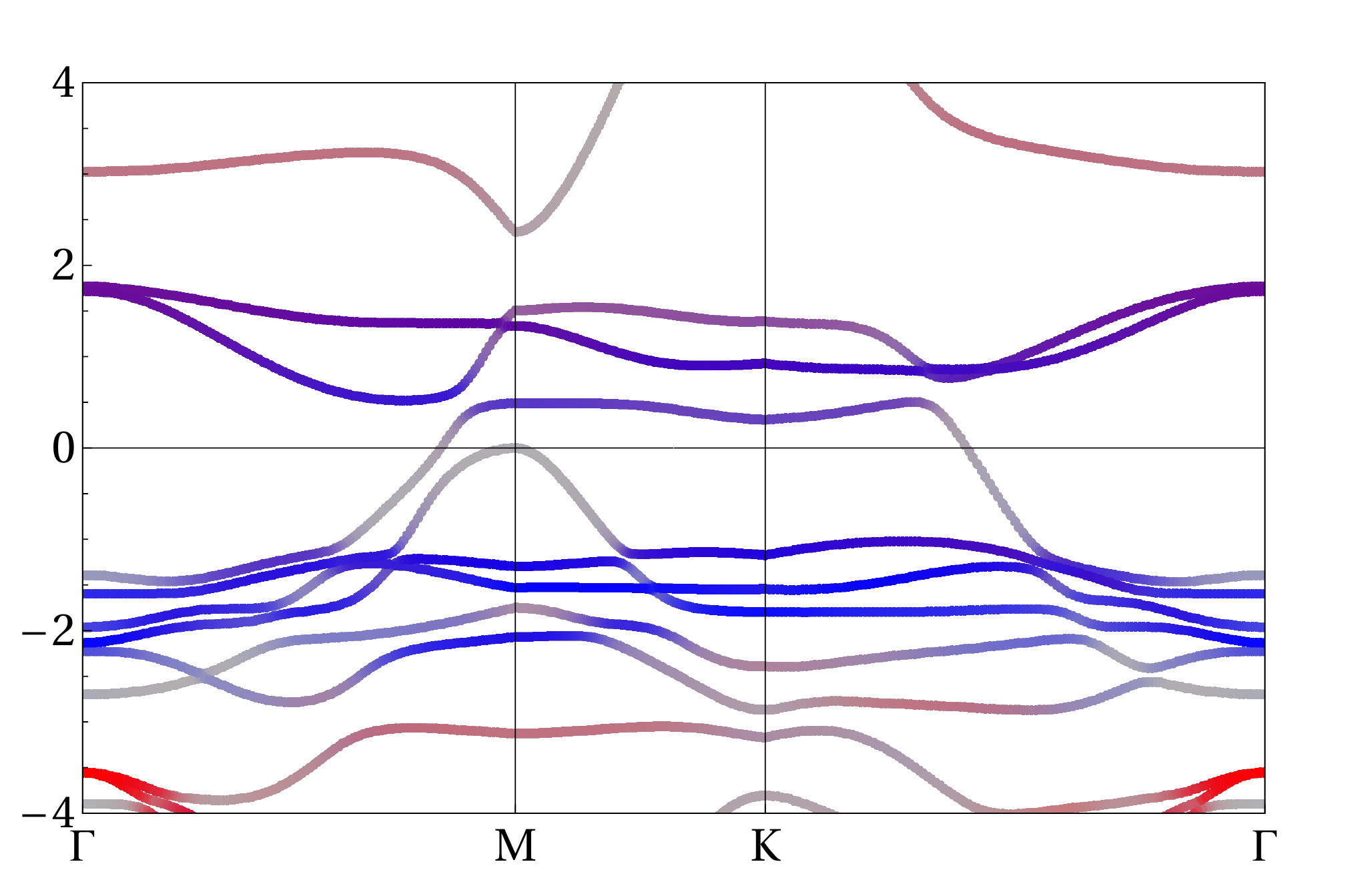}
\caption{
\label{fig:suppl:3d} Band structure of the twenty-band tight-binding model with spin-orbit coupling along high-symmetry lines (of the 2D Brillouin zone).
Weight of Co (O) electrons is represented by blue (red).
}
\end{figure*}

\section{$\mathbb{Z}_2$ Fu-Kane-Mele class and $\mathbb{Z}$ mirror Chern class}
\label{Supp:topology}

PtCoO$_2$ belongs to space group (SG) $R\bar{3}m$ (\#166) that is a symmorphic space group obtained from 
the combination of the rhombohedral Bravais lattice with point group $D_{3d} = \{ E, 2C_{3z},3C_2^\prime,I, 2IC_{3z},3 IC_2^\prime\}$. 
Due to TRS ($\mathcal{T}$) and inversion symmetry ($I$) every band is at least doubly degenerate, i.e. they form Kramers pairs, 
over the whole Brillouin zone. 
We show in Supplementary~Fig.~\ref{fig:suppl:bs}a 
the band structure of the 9-band tight-binding model of PtCoO$_2$, where we label the bands (Kramers pairs) from below as 
$\{\mathcal{B}_1,\mathcal{B}_2,\mathcal{B}_3,\mathcal{B}_4,\mathcal{B}_5,\mathcal{B}_6,\mathcal{B}_7,\mathcal{B}_8,\mathcal{B}_9\}$. 
We observe that every two successive bands are disconnected by a gap such that the topology of each band can be characterized separately. 
Importantly for the topological characterization of the bands we define that two bands are separated by a gap 
if we can find an energy gap between the bands at every $k$ of the Brillouin zone independently of the dispersion relation of the gap.
In the following we refer to the gap between the two successive bands 
$\mathcal{B}_i$ and $\mathcal{B}_{i+1}$ as $\Delta_{i,i+1}$. 
We note that the Fermi level crosses the band $\mathcal{B}_5$ and hence crosses the gap $\Delta_{45}$ and $\Delta_{56}$.

\begin{figure*}[t]
\centering 
\includegraphics[width=.7\hsize]{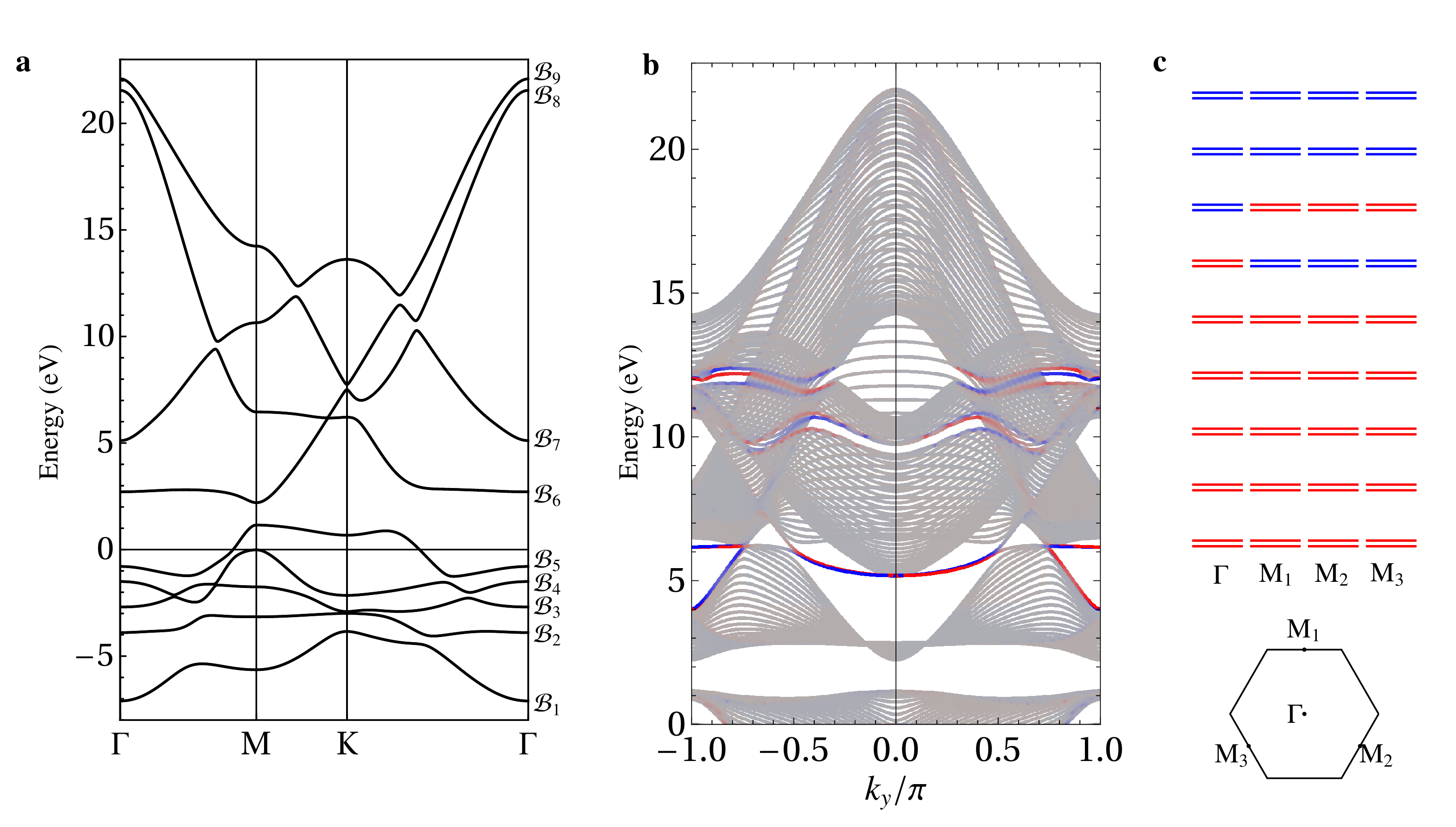}
\caption{
\textbf{a}, Band structure of the nine-band tight-binding model with spin-orbit coupling along high-symmetry lines. 
\textbf{b}, Band structure with edge states above the Fermi level. See also Fig.~\ref{fig:topo}d.
\textbf{c}, Inversion eigenvalues (red: $+1$, blue: $-1$) at the time-reversal invariant momenta, $\Gamma$, M$_1$, M$_2$, and M$_3$.
}
\label{fig:suppl:bs}
\end{figure*}

We show
the edge spectrum of the 9-band tight-binding model of PtCoO$_2$ obtained on a slab geometry,
 in Fig.~\ref{fig:topo}d and Supplementary~Fig.~\ref{fig:suppl:bs}b for below and above the Fermi level, respectively. 
We deduce the presence of edge states within the gaps $\{\Delta_{12},\Delta_{34},\Delta_{45},\Delta_{67},\Delta_{89}\}$, 
and no edge states within the gaps $\{\Delta_{23},\Delta_{56},\Delta_{78}\}$. 
In the following we determine their topological origin which turns out to be intrinsically related to the crystalline symmetries PtCoO$_2$. 

Centrosymmetry allows for a straightforward computation of the Fu-Kane-Mele $\mathbb{Z}_2$ invariant of every band 
in terms of the inversion eigenvalues at the time-reversal invariant momenta~\cite{FuKane_inversion}, see Supplementary~Fig.~\ref{fig:suppl:bs}c.
Supplementary~Figure~\ref{fig:suppl:bs}c shows that two bands above the Fermi level have nontrivial Fu-Kane-Mele topology ($\mathcal{B}_6$ and $\mathcal{B}_7$), 
while the Fu-Kane-Mele invariants of all the other bands are trivial. 
We therefore conclude that the edges states within the gap $\Delta_{67}$ are 
the helical edge states of a topological insulator which are protected by TRS only. 
The many other edge states observed in Fig.~\ref{fig:topo}d and Supplementary~Fig.~\ref{fig:suppl:bs}b, especially at the Fermi level, must have a different explanation.

It turns out that the bands of PtCoO$_2$ around the Fermi level only slightly break basal-mirror symmetry $\sigma_h$ (=$ I C_{2z}$) 
such that they bear features reminiscent of the higher symmetric space group SG191 whose point group is $D_{6h} = D_{3d} \times \{E,\sigma_h\}$. 
By globally imposing $\sigma_h$ symmetry the mirror Chern number~\cite{Teo_mirror_Z2} of every band is well defined by its presence. 
For this we simply compute the Chern number within each mirror symmetry sectors, i.e. $C_{1}^{+}$ ($C_{1}^{-}$) from even (odd) eigenstates under $\sigma_h$.
Practically it is most easily done by writing the Bloch Hamiltonian in the form $\mathcal{H} = \mathcal{H}_+ \oplus \mathcal{H}_-$, 
i.e. splitting the two mirror sectors, and then to compute the Chern number from a single block; 
the Chern number of the other block is then simply given through time reversal, i.e. $C_1^- = - C_1^{+}$. 
Remarkably we find a nonzero mirror Chern number in each band, namely $C_1^+(\mathcal{B}_1) = +2$, $C_1^+(\mathcal{B}_2) = -2$, 
$C_1^+(\mathcal{B}_3) = -6$, $C_1^+(\mathcal{B}_4) = +2$, $C_1^+(\mathcal{B}_5) = +4$, $C_1^+(\mathcal{B}_6) = -3$, $C_1^+(\mathcal{B}_7) = +3$, 
$C_1^+(\mathcal{B}_8) = -2$, $C_1^+(\mathcal{B}_9) = +2$, see Supplementary~Fig.~\ref{fig:suppl:wl}. 
We note that this agrees with the $\mathbb{Z}$ classification by K-theory of two-dimensional mirror symmetric topological insulators of the class AII~\cite{Sato_1}. 

It is now straightforward to deduce the edge states making use of the bulk-boundary correspondence for the Chern class~\cite{GrafPorta_BBC} within a single mirror sector, 
i.e the number of edge states is reflected in the difference in Chern number between gapped sets of bands. 
From the bulk mirror Chern numbers obtained above, 
we deduce the existence of two chiral edge branches crossing the gap $\Delta_{12}$, 
no edge states within the gap $\Delta_{23}$, 
six chiral edge branches crossing the $\Delta_{34}$, 
four chiral edge branches within $\Delta_{45}$, 
no edge states within the gap $\Delta_{56}$, 
three chiral edge branches within $\Delta_{67}$,
no edge states within the gap $\Delta_{78}$, and
two chiral edge branches within $\Delta_{89}$.
This seems to match perfectly with the edge spectrum of Fig.~\ref{fig:topo}d and Supplementary~Fig.~\ref{fig:suppl:bs}b, albeit sometimes hard to see due to the narrow band gaps.
Since the total mirror Chern number for the bands 1, 2, $\dots$, 5 is zero---the Chern class is additive in the vector bundle sense, 
i.e. $C_1^+(\mathcal{B}_1\oplus \mathcal{B}_2\oplus\dots\oplus\mathcal{B}_5) = C_1^+(\mathcal{B}_1)+C_1^+(\mathcal{B}_2)+\dots+C_1^+(\mathcal{B}_5)=+2-2-6+2+4=0$---we do not expect edge states within the gap $\Delta_{56}$, in agreement with Fig.~\ref{fig:topo}d. 
The gap $\Delta_{45}$ hosts chiral edge states that cross the Fermi level, as a consequence of the bulk mirror Chern numbers 
$C_1^+(\mathcal{B}_1\oplus\dots\oplus\mathcal{B}_4) = -4$ and $C_1^+(\mathcal{B}_5\oplus\dots\oplus\mathcal{B}_9) = +4$.

We conclude that, in addition to the two bands with nonzero Fu-Kane-Mele invariant, the edge states of PtCoO$_2$ are reminiscent of the nontrivial topology of the $\sigma_h$-symmetric parent model. 
Once again, we emphasize that this is because of the weak breaking of the basal mirror, giving the possibility to observe vicinity topologies as a results of approximate symmetries. Indeed, the robustness of the edge states at the Fermi level originates from the almost perfect $\sigma_h$ symmetry of the bands in the vicinity of the Fermi level, see Supplementary~Fig.~\ref{fig:suppl:EY}. 

\begin{figure*}[t]
\centering 
\includegraphics[width=\hsize]{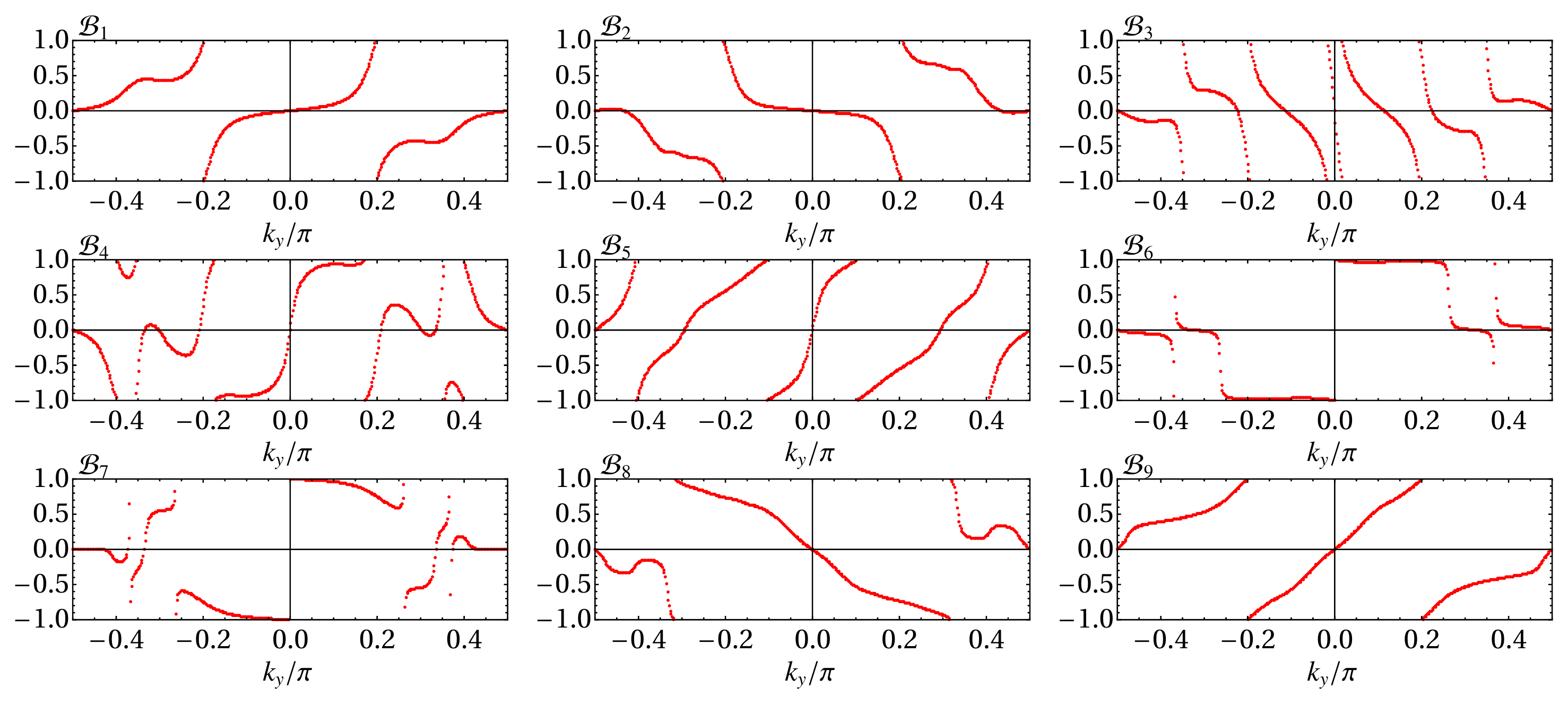}
\caption{
Flow of Wilson loop eigenvalues $X(k_y)$ (in unit of $a/2$) for the $\sigma_h=+i$ sector of the bands $\mathcal{B}_1$ up till and including $ \mathcal{B}_9$, for the mirror-symmetric Hamiltonian $H_{\mathcal{M}}$.
\label{fig:suppl:wl}
}
\end{figure*}

\section{Metastable (fragile) topology in PtCoO$_2$}
\label{sec:fragile}

We here want to briefly consider the connection to the recently coined concepts of metastable (fragile) topologies. 
It has been shown that hexagonal lattices naturally hosts nontrivial topology enforced by the crystal symmetries and TRS~\cite{BouhonSlager_1},
which can be related to the separation of elementary band representations by energy gaps~\cite{TQC,Cano_BEBRs}. 
The existence of separable elementary band representations for a given space group depends 
the Wyckoff positions occupied by the different atoms of the unit cell~\cite{TQC,Cano_BEBRs}. 
PtCoO$_2$ is made of layers of triangular lattices of Pt and Co, corresponding to the Wyckoff positions $3a$ and $3b$, 
respectively, and with, above and below each Pt atom, two O atoms corresponding to the Wyckoff position $6c$, 
none of which gives rise to a separable elementary representation~\cite{Elcoro:ks5574}. 

On the other hand, it has been shown in Ref.~\onlinecite{BouhonSlager_1} [see also Ref.~\onlinecite{Bradlyn_WilC3,1808.05375}] that $C_{2z}$ symmetry combined with TRS 
leads to a $\mathbb{Z}$ quantisation of Wilson loop windings. 
Indeed, the combined anti-unitary $C_{2z}\mathcal{T}$ symmetry, where $(C_{2z}\mathcal{T})^2=+1$, 
imposes that the the Wilson loop matrix of any Kramers pair of bands is $\mathcal{W}[l]\in \text{SO}(2)$. 
The classification of Wilson loop windings of a Kramers pair over the Brillouin zone is then readily given through $\pi_1(\text{SO}(2))=\mathbb{Z}$. 
Therefore, when $\sigma_h$ and $I$ are broken, while $C_{2z}$ is preserved, 
the Wilson loop windings of type of Fig.~\ref{fig:topo}c will remain, 
as a consequence of the fact that Wilson crossings at $0$ and $\pm \pi$ are protected by $C_{2z}\mathcal{T}$. 
This feature is stable for any Kramers pair of bands that is separated by an energy gap above and below from the other bands. 
However the fact that $\pi_1(\text{SO}(n))=\mathbb{Z}_2$ when $n>2$ readily tells us that 
it cannot pertain to thermodynamical (stable) properties of the system and rather characterises its metastable (fragile) properties. 
We thus conclude that these notions may find a suitable testbed in PtCoO$_2$.

\section{Spin polarization of the mirror eigenstate and the Elliott-Yafet mechanism for spin relaxation}
\label{sec:Elliott-Yafet}
In the Elliott-Yafet mechanism, the spin flipping process occurs even in nonmagnetic scattering 
because spin is no longer conserved in the presence of the SOC. 
The breaking of the spin conservation is especially crucial 
when the SOC is strong as in the present system. 
However, since the spin flip also alters the mirror eigenvalue, 
such processes are strongly suppressed when the mirror symmetry $\sigma_{h}$ is present.

To understand this in a more quantitative manner, 
let us see how much the spin and mirror conservation are broken in the Fermi surface 
of the present system. 
We denote the Kramers-degenerate eigenstates on the Fermi surface as
$|\theta\tilde{\sigma}\rangle$, where $\theta$ is the angle in momentum space and $\tilde{\sigma}=\tilde{\uparrow},\tilde{\downarrow}$ is defined 
to satisfy $\langle\theta\tilde{\uparrow}|\sigma_{z}|\theta\tilde{\downarrow}\rangle=0$. 
While these eigenstates are usual spin-polarized states when the SOC is absent, 
they are represented as
\begin{align}
|\theta\tilde{\uparrow}\rangle & =\sum_{m}[a_{\theta m}|m\rangle\otimes|\uparrow\rangle+b_{\theta m}|m\rangle\otimes|\downarrow\rangle],\label{eq:EY1}\\
|\theta\tilde{\downarrow}\rangle & =\sum_{m}[a_{\theta m}^{\ast}|m\rangle\otimes|\downarrow\rangle-b_{\theta m}^{\ast}|m\rangle\otimes|\uparrow\rangle],\label{eq:EY2}
\end{align}
with nonzero spin mixing $\sum_m|b_{\theta m}|^2$ due to the broken spin conservation with finite SOC.
Here $m$ labels the orbitals and $|\uparrow\rangle,|\downarrow\rangle$ is the spin basis.
This mixing leads to spin flip even in nonmagnetic scattering, 
whose amplitude is characterized by a factor 
$|\langle \theta^\prime\tilde{\uparrow}|W|\theta \tilde{\downarrow}\rangle|^2$
in the golden rule calculation ($W$: scattering matrix).

\begin{figure*}[t]
\centering 
\includegraphics[width=0.6\hsize]{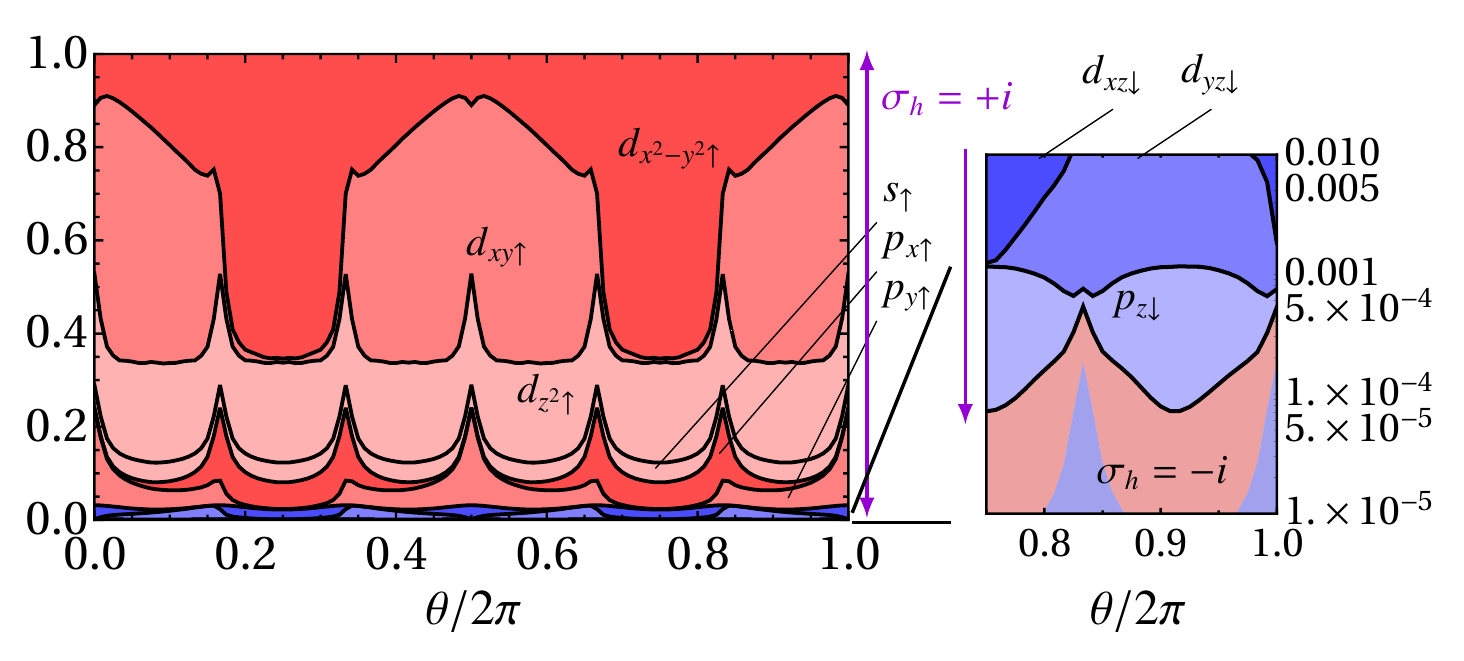}
\caption{
 Weight of spin and orbital component for the effective spin-up wave function on the Fermi surface $|\theta\tilde{\uparrow}\rangle$, 
parametrized by the angle in momentum space $\theta$. $\theta=0$ is set to the $\Gamma$--M direction. Red (blue) indicates spin-up (down) component.
Note the logarithmic scale in the zoomed plot in the right panel.
}
\label{fig:suppl:EY}
\end{figure*}

The amplitudes $|a_{\theta m}|^2$, $|b_{\theta m}|^2$ are plotted in Supplementary~Fig.~\ref{fig:suppl:EY} 
for $|\theta\tilde{\uparrow}\rangle$.
While the spin mixing of $\sum_{m}|b_{\theta m}|^{2}\sim0.03$ is as large as typical heavy elements (e.g., $\sim0.01$ for In, $\sim0.05$ for Pb, and $\sim0.1$ for Hg~\cite{Monod1979}),
the Fermi surface also has a rich orbital structure.
This may be advantageous since the spin flip amplitude also depends how the scattering matrix $W$ mixes the orbital degree of freedom.
In particular, when $W$ is mirror symmetric (such as in-plane defects of Pt atoms),
mirror (and spin) flip should be strongly prevented
thanks to the approximate mirror symmetry $\sigma_h$.
The mirror flipping amplitude is associated with the breaking of the mirror 
conservation, which is, as clearly shown in Supplementary~Fig.~\ref{fig:suppl:EY},
remarkably small $\sim10^{-4}$.
Namely, mirror-symmetric channels are much less effective on the spin relaxation,
and only mirror-breaking channels (such as oxygen vacancies) 
contribute to the spin relaxation.
Hence, the spin diffusion length in the present system is expected to be long in spite of a strong SOC.

\end{widetext}
\end{document}